\documentstyle[preprint,epsf,eqsecnum,aps,floats]{revtex}
\tightenlines
\begin{document}

\title{Search for High Mass Photon Pairs in 
$p \overline p \to \gamma\gamma jj$ Events at $\sqrt{s}=1.8
\rm~TeV$}

%
\author{                                                                      
B.~Abbott,$^{40}$                                                             
M.~Abolins,$^{37}$                                                            
V.~Abramov,$^{15}$                                                            
B.S.~Acharya,$^{8}$                                                           
I.~Adam,$^{39}$                                                               
D.L.~Adams,$^{49}$                                                            
M.~Adams,$^{24}$                                                              
S.~Ahn,$^{23}$                                                                
G.A.~Alves,$^{2}$                                                             
N.~Amos,$^{36}$
% ...
\\                                                                            
\vskip 0.70cm                                                                 
\centerline{(D\O\ Collaboration)}                                             
\vskip 0.70cm                                                                 
}                                                                             
\address{                                                                     
\centerline{$^{1}$Universidad de Buenos Aires, Buenos Aires, Argentina}       
\centerline{$^{2}$LAFEX, Centro Brasileiro de Pesquisas F{\'\i}sicas,         
                  Rio de Janeiro, Brazil}                                     
\centerline{$^{3}$Universidade do Estado do Rio de Janeiro,                   
                  Rio de Janeiro, Brazil}                                     
\centerline{$^{4}$Universidad de los Andes, Bogot\'{a}, Colombia}             
\centerline{$^{5}$DAPNIA/Service de Physique des Particules, CEA, Saclay,     
                  France}                                                     
\centerline{$^{6}$Panjab University, Chandigarh, India}                       
\centerline{$^{7}$Delhi University, Delhi, India}                             
\centerline{$^{8}$Tata Institute of Fundamental Research, Mumbai, India}      
\centerline{$^{9}$Kyungsung University, Pusan, Korea}                         
\centerline{$^{10}$Seoul National University, Seoul, Korea}                   
\centerline{$^{11}$CINVESTAV, Mexico City, Mexico}                            
\centerline{$^{12}$Institute of Nuclear Physics, Krak\'ow, Poland}            
\centerline{$^{13}$Institute for Theoretical and Experimental Physics,        
                   Moscow, Russia}                                            
\centerline{$^{14}$Moscow State University, Moscow, Russia}                   
\centerline{$^{15}$Institute for High Energy Physics, Protvino, Russia}       
\centerline{$^{16}$University of Arizona, Tucson, Arizona 85721}              
\centerline{$^{17}$Lawrence Berkeley National Laboratory and University of    
                   California, Berkeley, California 94720}                    
\centerline{$^{18}$University of California, Davis, California 95616}         
\centerline{$^{19}$University of California, Irvine, California 92697}        
\centerline{$^{20}$University of California, Riverside, California 92521}     
\centerline{$^{21}$Florida State University, Tallahassee, Florida 32306}      
\centerline{$^{22}$University of Hawaii, Honolulu, Hawaii 96822}              
\centerline{$^{23}$Fermi National Accelerator Laboratory, Batavia,            
                   Illinois 60510}                                            
\centerline{$^{24}$University of Illinois at Chicago, Chicago,                
                   Illinois 60607}                                            
\centerline{$^{25}$Northern Illinois University, DeKalb, Illinois 60115}      
\centerline{$^{26}$Northwestern University, Evanston, Illinois 60208}         
\centerline{$^{27}$Indiana University, Bloomington, Indiana 47405}            
\centerline{$^{28}$University of Notre Dame, Notre Dame, Indiana 46556}       
\centerline{$^{29}$Purdue University, West Lafayette, Indiana 47907}          
\centerline{$^{30}$Iowa State University, Ames, Iowa 50011}                   
\centerline{$^{31}$University of Kansas, Lawrence, Kansas 66045}              
\centerline{$^{32}$Louisiana Tech University, Ruston, Louisiana 71272}        
\centerline{$^{33}$University of Maryland, College Park, Maryland 20742}      
\centerline{$^{34}$Boston University, Boston, Massachusetts 02215}            
\centerline{$^{35}$Northeastern University, Boston, Massachusetts 02115}      
\centerline{$^{36}$University of Michigan, Ann Arbor, Michigan 48109}         
\centerline{$^{37}$Michigan State University, East Lansing, Michigan 48824}   
\centerline{$^{38}$University of Nebraska, Lincoln, Nebraska 68588}           
\centerline{$^{39}$Columbia University, New York, New York 10027}             
\centerline{$^{40}$New York University, New York, New York 10003}             
\centerline{$^{41}$University of Rochester, Rochester, New York 14627}        
\centerline{$^{42}$State University of New York, Stony Brook,                 
                   New York 11794}                                            
\centerline{$^{43}$Brookhaven National Laboratory, Upton, New York 11973}     
\centerline{$^{44}$Langston University, Langston, Oklahoma 73050}             
\centerline{$^{45}$University of Oklahoma, Norman, Oklahoma 73019}            
\centerline{$^{46}$Brown University, Providence, Rhode Island 02912}          
\centerline{$^{47}$University of Texas, Arlington, Texas 76019}               
\centerline{$^{48}$Texas A\&M University, College Station, Texas 77843}       
\centerline{$^{49}$Rice University, Houston, Texas 77005}                     
}                                                                             

\maketitle

\begin{abstract}
    A search has been carried out for events in the channel $p \overline p \to
\gamma\gamma jj$.
Such a signature can characterize the production of a non-standard 
Higgs boson together with a $W$ or $Z$ boson.
We refer to this non-standard Higgs, having standard model
couplings to vector bosons but no coupling to fermions, as a
``bosonic Higgs.''
With the requirement of two high transverse energy photons and two jets, 
the diphoton mass ($m_{\gamma\gamma}$) distribution is
consistent with expected background.  A 90(95)\% C.L.
upper limit on the cross section as a function of mass is calculated, ranging
from 0.60(0.80) pb for $m_{\gamma\gamma} = 65$ GeV/$c^2$ to 0.26(0.34) pb
for $m_{\gamma\gamma} = 150$ GeV/$c^2$, corresponding to a 95\% C.L. lower 
limit on the mass of a bosonic Higgs of $ 78.5$ GeV/$c^{2}$.
       
\end{abstract}

\newpage

The Higgs sector of the standard model is poorly
constrained.  Several extended Higgs models
~\cite{Akeroyd1,Akeroyd2,Haber,Weiler,Bamert,Georgi} allow
a light neutral scalar Higgs with suppressed couplings to
fermions.  We refer to such a non-standard Higgs, having 
standard model 
couplings to vector bosons but zero couplings to fermions, as a
``bosonic Higgs.'' The model of Refs.~\cite{Akeroyd1,Akeroyd2}
provides a bosonic Higgs without requiring fine tuning
and maintains the relation 
$\rho= M^{2}_{W}/M^{2}_{Z}\cos^{2}\theta_{W} = 1$,
consistent with present experimental limits\cite{Review}.

\begin{figure}[t]
\vspace{-1.cm}
\epsfxsize=9cm
\vskip.2in\centerline{\epsfbox{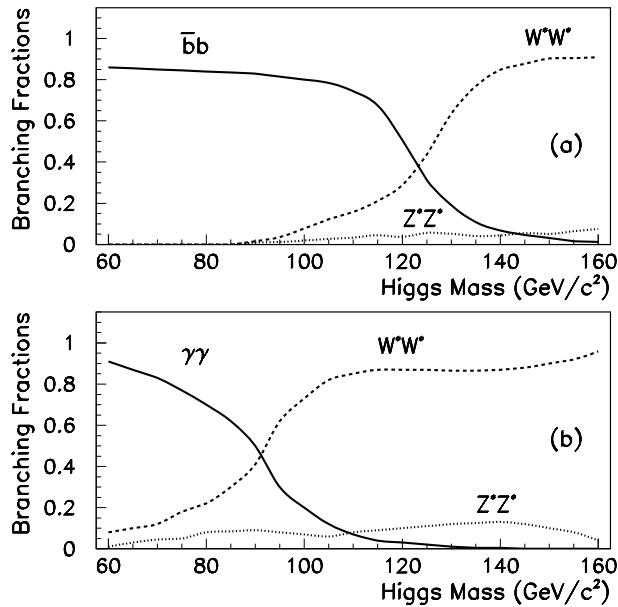}}
\caption{
Decay branching fractions vs. Higgs mass for (a) 
standard model Higgs, and (b) bosonic Higgs.  In (a), 
the diphoton branching
fraction is less than 0.001; $c\bar{c}$ and $\tau^+\tau^-$ Higgs decay
channels are not shown.  In (b), the Higgs decays to only VV, where
V = $\gamma$,~$W$, or~$Z$.
There is a large enhancement in the diphoton channel for the
bosonic Higgs model: the absence of competing decay channels results
in a dominant $H \rightarrow \gamma\gamma$ below 
$M_H \approx 90$ GeV/$c^2$.
\label{branch}
}
\end{figure}

The decay channels of a bosonic Higgs differ from those of
the standard model Higgs as shown in Fig.~\ref{branch}.
Since the fermion decay channels are suppressed,
the decay of a bosonic Higgs with mass less than $2M_W$ is not
dominated by $H \to b\bar{b}$.
At tree level the bosonic Higgs decays only to $WW^{(*)}$ and
$ ZZ^{(*)}$ vector bosons (where the asterisks denote that one or both
of the vector bosons may be off the mass shell).
For bosonic Higgs masses less than 90~\rm GeV/$c^2$,  
the one-loop $W$-boson-mediated $ H\to\gamma\gamma$ channel 
becomes dominant.

A bosonic Higgs is most easily detected in the associated production mode,
where an off-mass-shell $W$ or $Z$ boson is produced and radiates a
Higgs boson \cite{Willenbrock}.
Higgs production through vector boson fusion 
also contributes to the $\gamma\gamma jj$ final state at the 15\% level.
The sum of the $WH$ and $ZH$ production cross 
sections ranges from $1.8~\rm pb$ for $ M_H=60~{\rm \rm GeV}/c^2$,
to $0.4~ \rm pb$ for $M_H=100~{\rm GeV}/c^2$.
We expect sensitivity in the $\gamma\gamma jj$ final state 
up to a mass of $ M_H \sim $ 85 GeV/$c^2$ for the 
decay modes $ H\to\gamma\gamma$ and $ W/Z \to jj$, at which mass
the branching ratio $ H\to\gamma\gamma$ remains high and the falling
Higgs production cross section of $\sim$ 0.4 pb allows the production 
of tens of events.
This Letter describes the first search for a bosonic Higgs at hadron
colliders.  


Experiments at the LEP collider have previously set lower mass limits on a
bosonic Higgs.  A limit of approximately $60$ GeV/$c^2$ was 
established\cite{Willenbrock,L3Coll} in data taken
at the $Z^0$, a higher 95\% C.L. limit set at $76.5$ GeV/c$^2$ 
in 172 GeV collisions\cite{OPAL-1} 
at LEP2, and this limit extended to $90.0$ GeV/c$^2$ in 183 GeV 
collisions.\cite{OPAL-2}

Data corresponding to an integrated luminosity of
$101.2\pm 5.5\,{\rm pb}^{-1}$, recorded during 1992--96 with the
D\O\ detector\cite{dzero}, are used for this analysis.
Photons and jets are identified using the uranium-liquid-argon
sampling calorimeter, extending to a pseudorapidity $|\eta| =
|-\ln\tan{\theta\over 2}|
\lesssim 4.5$, where $\theta$ is the polar angle. 
The electromagnetic (EM) energy resolution is
$\sigma_E/{E} = 15\%/\sqrt{E (\rm GeV)} \oplus 0.3\%$, and the
jet energy resolution is about $\sigma_E/{E} = 80\%/\sqrt{E
(\rm GeV)}$. The calorimeter
is segmented transversely 
into towers in pseudorapidity and azimuthal angle of size
$\Delta \eta \times \Delta \phi = 0.1 \times 0.1$, and further
segmented to $0.05 \times 0.05$ at the EM shower maximum.
Drift chambers in front of the calorimeter are used to distinguish
photons from electrons.
A three-level triggering system is employed:
level 0 uses scintillation counters near the
beam pipe at each end of the detector to detect an inelastic interaction;
level 1 sums the EM and hadronic energy in 
calorimeter towers of size
$\Delta \eta \times \Delta \phi =0.2 \times 0.2 $; and
level 2 is a software
trigger which forms clusters of calorimeter cells and applies
loose cuts on the shower shape.

Events used in this analysis have at least two photon candidates and
at least two jet candidates.  Initially, the events are selected using
a diphoton trigger that requires two EM showers with a transverse
energy ($E_T$) greater than 12 GeV.  The filter is fully efficient
when both photons have $E_T > 15$ GeV.  The offline event selection
criteria are optimized by requiring one photon to have
$\mbox{$E_T^\gamma$} > 30\,$\rm GeV and $|\mbox{$\eta^\gamma$}|<1.1$
or $1.5<|\mbox{$\eta^\gamma$}|<2.0$, and the other to have
$\mbox{$E_T^\gamma$} > 15\,$GeV and $|\mbox{$\eta^\gamma$}|<1.1$ or
$1.5<|\mbox{$\eta^\gamma$}|<2.25$.  Additionally, one hadronic jet is
required to have $\mbox{$E_T^{\rm jet}$} > 30\,$GeV and
$|\mbox{$\eta^{\rm jet}$}|<2.0$, and the other hadronic jet to have
$\mbox{$E_T^{\rm jet}$} > 15\,$GeV and $|\mbox{$\eta^{\rm
jet}$}|<2.25$.  For the two jets to be consistent with the decay of a
$W$ or $Z$ boson, the dijet mass is required to be between 40
GeV/$c^2$ and 150 GeV/$c^2$. A photon candidate is rejected if there
is either a reconstructed track or a significant number of drift
chamber hits in a tracking road $\Delta\theta\times\Delta\phi =
0.2\times 0.2$ between the cluster in the calorimeter and the
interaction vertex. A photon candidate is required to have a shower
shape consistent with that of a single EM shower, to have more than
96\% of its energy in the EM section of the calorimeter, and to be
isolated \cite{prl}.  Isolation requires that the transverse energy in
the annular region between ${\cal R} \equiv \sqrt{\Delta\eta^2 +
\Delta\phi^2}=0.2$ and ${\cal R}=0.4$ around the cluster be less than
10\% of the total cluster transverse energy.  In addition, each photon
candidate must be separated by $\Delta\cal{R}_{\gamma} >$ 0.7 from
every jet~\cite{Lauer-thesis}.  Each jet candidate is reconstructed
from energy deposited in a $\Delta\cal{R} <$ 0.7 cone, must have less
than 95\% of its energy in the EM section of the calorimeter, and must
have no more than 40\% of its energy in the outermost layer of the
hadronic portion of the calorimeter.

\begin{figure}[t]
\vspace{-1.cm}
\epsfxsize=9cm
\vskip.2in\centerline{\epsfbox{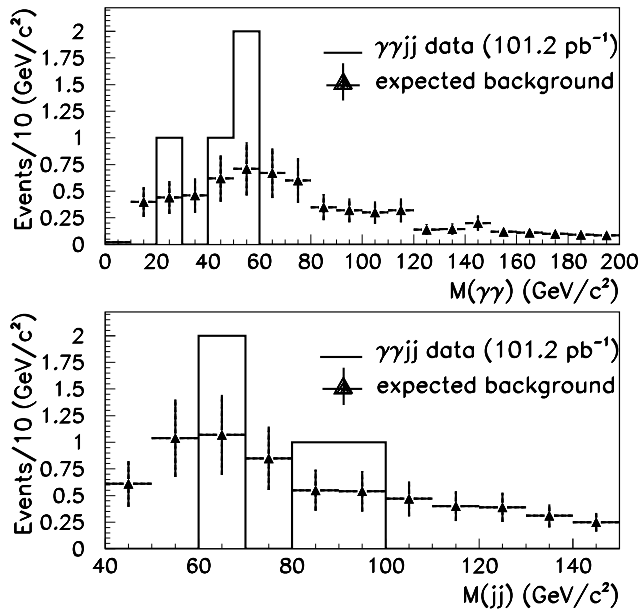}}
\caption{The data and expected background for (a) the diphoton
mass and (b) the dijet mass distributions.  
\label{bh_signback_fincuts}
}
\end{figure}

These selection criteria yield four events, whose $m_{\gamma\gamma}$ 
distribution is shown in Fig.~\ref{bh_signback_fincuts}a.  
No events are observed with $ m_{\gamma\gamma} > $ 60 GeV/$c^2$.
The resolution of the detector in $ m_{\gamma\gamma}$ is about
2.5 GeV/$c^2$ for diphoton final states passing these kinematic cuts.
The corresponding dijet mass distributions of data and expected background
are shown in Fig.~\ref{bh_signback_fincuts}b.

The dominant background to the $\gamma\gamma jj$ channel is production
of QCD multijet
events in which two jets are misidentified as photons.
During the jet fragmentation process, $\pi^{0}$ and
$\eta$ mesons are produced and decay promptly into multiple photons.
If the $\pi^{0}$ or $\eta$ meson carries a large fraction of the 
jet energy and has a momentum greater than about 
10 GeV/$c$ the decay photons 
coalesce to mimic a single isolated photon in the calorimeter.  
The depth development of multiple photons differs from that of a
single photon, and a fit to the longitudinal shower shape for
photon candidates yields
the probability $P(j \rightarrow ``\gamma$'') for a jet to mimic an 
isolated photon candidate, estimated to be 
$(4.3\pm 1.0) \times 10^{-4}$, with a weak $E_T$ 
dependence~\cite{Lauer-thesis}.

Smaller sources of background are from
double direct photon production,
single direct photon production with one jet fluctuating into a
photon candidate, and final states containing electrons 
in addition to two jets, such as
~($W \to e^{\pm}\nu)\gamma jj$, 
($Z \to e^+e^-) jj$,  and 
$t\bar{t} \to e^+e^-\nu\nu jj$, where the electrons fail track
reconstruction and are misidentified as photons.  
By rejecting events that have a track or significant number of
drift chamber hits inside the tracking road, the expected electron 
background is reduced to 
less than 0.01 events, and is not considered further.  

The QCD multijet background to $\gamma\gamma jj$ events is estimated
from the data.  
Starting with the same trigger and data set as the signal sample,
a background sample is selected by requiring two EM clusters and jets
satisfying the same kinematic and fiducial criteria as the signal. 
Both EM clusters are required to
have more than 90\% of their energy in the EM section of the calorimeter
and to have no reconstructed track associated with the cluster,
but at least one of the two EM clusters is required to fail the photon
quality criteria (isolation, shower shape, or EM fraction).
The resulting sample of 194 events is expected to be dominated by 
QCD multijet events where two jets
fluctuate into highly-electromagnetic clusters.  
After
subtracting the expected direct photon event contribution, 
the QCD multijet background for $ m_{\gamma\gamma} > $60 GeV/$c^2$ 
is estimated by normalizing the cluster-pair mass distribution to the
signal sample over the mass range $ m_{\gamma\gamma} < $60 GeV/$c^2$,
where bosonic Higgs have been excluded by 
earlier searches for $Z\to Z^*H$ at LEP\cite{L3Coll,OPAL-1}. 

The direct photon background is calculated 
using the {\sc pythia} Monte Carlo\cite{pythia}.  
This background has two sources: single direct
photon production where one true photon is produced and one 
jet is misidentified as a photon, and double direct
photon production, where two true photons are produced in
addition to two high-$E_T$ jets.  
The Monte Carlo jet and photon energies are smeared to match the 
measured detector resolutions.  The efficiency for the 
events to pass the photon quality criteria
(isolation, shower shape, EM fraction, and tracking) are
calculated from data 
using our $ Z \to e^+e^-$ event sample.  The 
single direct photon events are weighted by the probability 
P($j \rightarrow ``\gamma$''), since one of
the jets must be misidentified as a photon for a background event 
to be accepted.  The direct photon background is
normalized to the signal sample using the calculated direct photon 
cross section.  The dominant systematic uncertainty in these sources
of background then derives from the observed
level of agreement between the theoretical and experimental direct photon
cross sections, and is estimated to be 40\% for double direct photon
production and 20\%
for single direct photon production~\cite{prl,chen}.  

Figure 2 shows the total expected background, with estimated
uncertainties, in bins of 10 GeV/$c^2$.  The total background of $6.0
\pm 1.6$ events consists of $4.0 \pm 1.5$ QCD multijet events and $2.0
\pm 0.6$ direct photon events.  It agrees well with our observed
number of four events.  We find no evidence for non-standard sources
of $\gamma\gamma jj$ events.  If we increase the photon pseudorapidity
coverage to $|\mbox{$\eta^\gamma$}|<2.5$, and reduce the leading jet
and photon transverse energy requirements to 15 GeV, the same
background estimation technique predicts $38\pm 10$ events while 39
events are observed.  The $m_{\gamma\gamma}$ and $m_{jj}$
distributions of this larger sample are also described well by the
estimated background.

Seven samples of 
bosonic Higgs events, each containing 
5000 simulated events, are generated using 
{\sc pythia} for 
the processes $p\bar{p} \rightarrow WH$ and $p\bar{p} \rightarrow ZH$,
with the decays $H \rightarrow \gamma \gamma$ and $W$/$Z$
$\rightarrow qq^{\prime}$, for Higgs masses of 60, 70, 80, 90, 100, 110,
and 150 GeV/$c^{2}$.  These events are processed through the 
detector simulation and event reconstruction software.  The Higgs
selection criteria 
are applied and the signal acceptance and efficiency calculated;
their product ranges from 0.06 to 0.10 for Higgs masses of 60 to 150 
GeV/$c^2$.  The systematic uncertainty in the 
acceptance for the Higgs signal is
based on the level of agreement between the Monte Carlo and 
data-based estimates of the photon and jet selection efficiencies.
The systematic error includes the uncertainties in the efficiencies
for the photon trigger and selection (2\%), requirement on
photon track rejection
(13\%), hadronic energy scale (5-11\%),
EM energy scale ($\simeq$1\%), and jet reconstruction 
($\simeq$1\%).  The statistical uncertainty on the Monte Carlo Higgs
samples is about 3\%.  The systematic and statistical uncertainties,
and the integrated luminosity uncertainty of 5.3\%, are added in 
quadrature and yield 15\%. 

\begin{figure}[t]
\vspace{-1.cm}
\epsfxsize=9cm
\vskip.2in\centerline{\epsfbox{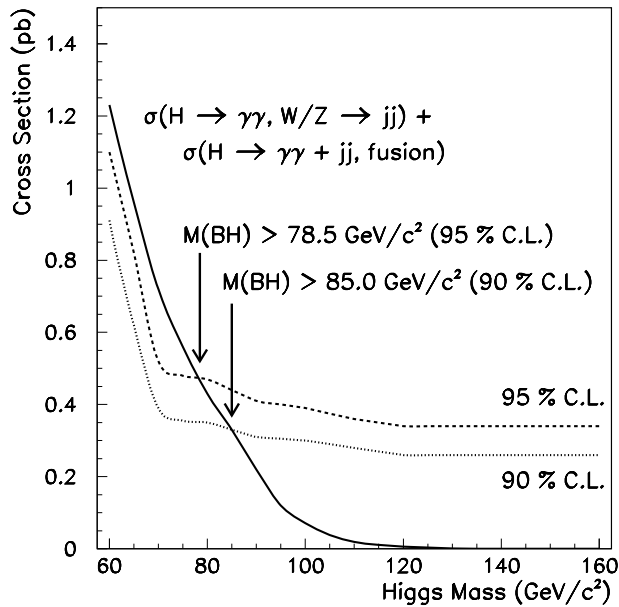}}
\caption{The solid curve represents the bosonic Higgs 
95\% C.L. exclusion contour, 
the dashed curve represents the 90\% C.L. exclusion contour, and
the dot-dash curve represents the bosonic Higgs cross section with
$H \to \gamma\gamma$ and $W/Z \to jj$ branching fractions taken into account.
\label{bh_9095limit}
}
\end{figure}

The 90\% and 95\% confidence
level (C.L.) limits on the cross section as a function of
$m_{\gamma \gamma}$ are shown in Fig.~\ref{bh_9095limit} 
and are calculated using
a Bayesian approach \cite{Review}, 
incorporating the uncertainties associated with the 
efficiency, acceptance, 
luminosity, and the expected background as a function of
$m_{\gamma\gamma}$ and $m_{jj}$.  
Correlations between errors 
are negligible and not included.
A general 95\% C.L. upper limit on the cross section is calculated 
from the
exclusion contour in Fig.~\ref{bh_9095limit}, and  ranges from
0.80 pb for $m_{\gamma\gamma} = $ 65 GeV/$c^2$ to 0.34 pb 
for $m_{\gamma\gamma} = $ 150 GeV/$c^2$.  The corresponding 90\% C.L.
is noted in the figure.
The full bosonic Higgs cross section is also plotted in the 
Fig.~\ref{bh_9095limit}, and includes both the associated
production and vector boson fusion production processes,
calculated using {\sc pythia} with a QCD correction 
factor\cite{Willenbrock}  of $1.25$.  This factor agrees with the
ratio between our measured cross section for W boson production\cite{D0-W}
and the calculated cross section, $1.23 \pm 0.08$.
We  set lower limits on the bosonic Higgs mass of 85.0 GeV/$c^2$ at the 
90\% C.L. and 78.5 GeV/$c^2$ at the 95\% C.L.

In summary, we performed the first search for a bosonic Higgs at hadron
colliders, in the channel 
$p\bar{p}\to\gamma\gamma jj$.   Four candidates pass the selection
requirements, with an expected background of  $6.0 \pm 2.1$ events.
No candidate events are seen with a diphoton
mass greater than 60 GeV/$c^2$. 
A 95\% C.L. bosonic Higgs lower mass limit of 78.5 GeV/$c^2$ 
is set, assuming
standard model couplings between the Higgs and the vector bosons.
The 95\% C.L. upper limits on the bosonic Higgs production cross section 
range from 0.80 $ \rm pb$ for $m_{\gamma\gamma} = $ 65 GeV/$c^2$ 
to 0.34 $ \rm pb$ for $m_{\gamma\gamma} = $ 150 GeV/$c^2$.
                                    
%
We thank the Fermilab and collaborating institution staffs for
contributions to this work and acknowledge support from the 
Department of Energy and National Science Foundation (USA),  
Commissariat  \` a L'Energie Atomique (France), 
Ministry for Science and Technology and Ministry for Atomic 
   Energy (Russia),
CAPES and CNPq (Brazil),
Departments of Atomic Energy and Science and Education (India),
Colciencias (Colombia),
CONACyT (Mexico),
Ministry of Education and KOSEF (Korea),
and CONICET and UBACyT (Argentina).


\vspace{-0.3cm}

\begin{references}

%
\bibitem[*]{ecuador}
Visitor from Universidad San Francisco de Quito, Quito, Ecuador.

\bibitem[\dag]{beijing}
Visitor from IHEP, Beijing, China.

\vskip 0.25cm

%

\vspace{0.25cm}

\bibitem{Akeroyd1} A.G. Akeroyd, Phys. Lett. B~ \bf 368\rm, 89 (1996).
%
\bibitem{Akeroyd2} A.G. Akeroyd, Phys. Lett. B~ \bf 353\rm, 519 (1995).
%
\bibitem{Haber} H.E. Haber, G.L. Kane and T. Sterling, Nucl. Phys. \bf B161\rm,
493 (1979).
%
\bibitem{Weiler} H. Pois, T. Weiler and T.C. Yuan, Phys. Rev. D \bf 47\rm,
 3886 (1993).
%
\bibitem{Bamert} P. Bamert and Z. Kunszt, Phys. Lett. B \bf 306\rm, 335 (1993).
%
\bibitem{Georgi} H. Georgi and M. Machacek, Nucl. Phys. \bf B262\rm,
 463 (1985); M. Chanowitz and M. Golden, Phys. Lett. {\bf 165B}, 105 (1985).
%
\bibitem{Review} R.M. Barnett \it et al.\rm, Phys. Rev. D \bf 54\rm, 1 (1996).
%
\bibitem{Willenbrock} A. Stange, W. Marciano and S. Willenbrock, Phys. Rev.
D {\bf 49},  1354 (1994).
%
\bibitem{L3Coll} L3 Collaboration, 
M. Acciarri {\it et al.}, Phys. Lett. B \bf 388\rm, 409
	(1996).
%
\bibitem{OPAL-1} OPAL Collaboration, 
K. Ackerstaff, {\it et al.}, Eur. Phys. J. {\bf C1} (1998) 31.
%
\bibitem{OPAL-2} OPAL Collaboration,
K. Ackerstaff, {\it et al.}, Phys. Lett. B {\bf 437}, 218 (1998).
%
\bibitem{dzero} D\O\ Collaboration, S. Abachi {\it et al.},
Nucl. Instrum. Methods Phys. Res. A {\bf 338} (1994) 185.
%
\bibitem{prl} D\O\ Collaboration,
S.~Abachi {\it et al.}, Phys. Rev. Lett. {\bf 78}, 2070 (1997).
%
\bibitem{Lauer-thesis} Bryan A. Lauer, Ph.D. thesis, Iowa State University,
Ames, IA (1997), available at {\tt
http://\-www-d0.fnal.gov/ {results/publications\_talks/thesis/lauer/thesis.ps}}
(unpublished).
%
\bibitem{pythia} T. Sjostrand and M. Bengtsson, Comput. Phys. Commun. 
{\bf 43}, 367 (1987). Version 5.7 was used in this analysis.
%
\bibitem{chen} W. Chen, Ph.D. Thesis, State University of New York, 
Stony Brook,
NY (1997), available at {\tt
http://\-www-d0. fnal\-.gov/\-results/\-publications\_talks/\-thesis/\-chen/
\-chen.html} (unpublished).
\bibitem{D0-W} D\O\ Collaboration,  S.~Abachi {\it et al.}, Phys. Rev. Lett.
{\bf 75}, 1456 (1995).
%
\end{references}
\end{document}